\pdfoutput=1
\documentclass[journal]{IEEEtran}

\usepackage{amsmath,amssymb,amsthm}
\usepackage{mathrsfs}
\usepackage{subfigure}
\usepackage{cite}
\usepackage{multirow}
\usepackage{array}
\usepackage{graphicx}
\usepackage{bm}
\usepackage[flushleft]{threeparttable}
\usepackage{algorithm2e}
\usepackage{algpseudocode}
\usepackage{booktabs}
\usepackage{xfrac}
\usepackage[table]{xcolor}
\usepackage{url}
\usepackage[hidelinks]{hyperref}

\newcommand{\subb}[1]{{\bm{#1}}}

\begin{document}
	
	\title{3D Radar Imaging from the UAV Nadir} %\textcolor{red}{H: A Dual-Layered Autofocus Strategy Integrating Polar Format and Backprojection Algorithms for Automotive SAR Imaging}}

\author{S.~Hamed~Javadi, Hichem~Sahli and~André~Bourdoux,~\IEEEmembership{Senior Member,~IEEE}
	\thanks{Accepted for publication in IEEE Transactions on Radar Systems. DOI: \href{https://doi.org/10.1109/TRS.2026.3731078}{10.1109/TRS.2026.3731078}.}
	\thanks{The authors are with Interuniversity Micro-Electronics Center (IMEC), Kapeldreef 75, B-3001 Leuven, Belgium. H. Sahli is also with Informatics Dept., Vrije Universiteit Brussel (VUB), Pleinlaan 2, 1050 Brussels, Belgium. (email: hamed.javadi@imec.be; hichem.sahli@imec.be; andre.bourdoux@imec.be).}
}
%% The paper headers
\markboth{Accepted manuscript, IEEE Transactions on Radar Systems}%
{Javadi, Sahli, and Bourdoux: 3D Radar Imaging from the UAV Nadir}

\maketitle
\begin{abstract}
	Radars improve the sensing robustness of UAVs by operating under poor lighting and weather conditions and seeing through occlusions such as vegetation. However, they suffer from poor angular resolution, which can be addressed using synthetic aperture radar (SAR) algorithms. State-of-the-art UAV SAR methods operate at a depression angle and are not suitable for sensor fusion applications where the data are collected from areas directly below the UAV (i.e., the UAV nadir). In this paper, we present an interferometric SAR (InSAR) framework for reconstructing 3D images from the UAV nadir using a low-cost multi-input-multi-output (MIMO) mm-wave radar. Additionally, an effective method based on the phase gradient autofocus (PGA) is presented for compensating the phase error across the virtual receive antennas. We demonstrate the effectiveness of our 3D imaging algorithm in both simulation and experimental scenarios.
\end{abstract}

\begin{IEEEkeywords}
	Frequency modulated continuous wave (FMCW), interferometric synthetic aperture radar (InSAR), multiple-input-multiple-output (MIMO), phase gradient autofocus (PGA), polar format algorithm (PFA), radar imaging, synthetic aperture radar (SAR), unmanned aerial vehicle (UAV).
\end{IEEEkeywords}

\section{Introduction}
\IEEEPARstart{U}{nmanned} aerial vehicles (UAVs), or drones, have become pervasive across diverse applications thanks to their easy deployment and low cost. Each application is enabled by employing a proper set of sensors on board. Radar is a powerful sensor that enhances the sensing robustness of drones by functioning in poor light and adverse weather conditions and seeing through occlusions such as fog and vegetation. Accordingly, there is a growing tendency to adopt it in civilian and military applications of UAVs.

The radars bottleneck is their requirement of a large antenna aperture size for achieving high angular resolution, which is either impractical or costly. Alternatively, the required aperture can be synthesized by leveraging the radar motion, giving rise to synthetic aperture radar (SAR) algorithms \cite{Richards2014fundamentals,Doerry_PFA}. 

The first practical SAR implementation dates back to 1978 for oceanographic observations \cite{Ausherman1984,Franceschetti1999}. Since then, numerous SAR processing algorithms---including range-Doppler (RD), polar format algorithm (PFA), and backprojection algorithm (BPA)---have been developed, each optimized for different trade-offs between computational complexity and image quality. These algorithms have primarily been deployed in airborne and spaceborne platforms. These systems typically operate at depression angles between approximately $20^\circ$ and $70^\circ$, depending on the required swath width and imaging resolution \cite{sarhandbook2019chapter5, Simard2016}. Subsequently, either multiple receive antennas (along the across-track direction) or multiple radar passes are used for 3D imaging by interferometric SAR (InSAR) \cite{Richards_InSAR_2007,Brenner2008,Eineder2009}.

Recent advances in GPU acceleration have enabled real-time onboard SAR processing for UAVs during flight rather than post-processing on the ground. Using a frequency modulated continuous wave (FMCW) radar in the mm-wave frequency range, Bekar et al. \cite{uav_sar_Bekar_2022} presented a drone-borne SAR system for short range imaging. The impact of the attitude angles (i.e., roll, pitch, and yaw) of the UAV on the SAR image quality was shown in \cite{Svedin_Bernland_Gustafsson_Claar_Luong_2021}.

The concept of InSAR has been adopted in \cite{tomoSAR_Wang_2024} for tomography purposes, where the authors address shadowing and layover by combining 3D structures captured from different perspectives and evaluate their algorithm in simulation. In \cite{Lahmeri2025}, two UAVs send their radar observation data to a ground processing unit for 3D imaging via InSAR.  Mustieles-Perez et al. \cite{Drone_InSAR_2025} outline the detailed steps of drone-borne InSAR.

Current UAV SAR approaches are designed primarily based on airborne SAR configurations, producing imagery in down-range and azimuth dimensions at a given depression angle. Thus, the state-of-the-art methods are unable to reconstruct imagery of the region directly beneath the UAV (its nadir). In contrast, UAV platforms typically carry downward-looking sensors such as RGB, multispectral, thermal, and LiDAR to capture data from the ground area directly below \cite{Kellner2019,Gano2024,Klemas2015}. To leverage radar imagery for sensor fusion, it is essential that radar systems also observe this nadir region. Furthermore, performing SAR at a depression angle demands a larger maximum unambiguous range compared to nadir SAR imaging, which limits the maximum UAV altitude.

Leveraging the SAR geometry presented in \cite{Javadi_depth_map_eurad2025,Javadi_altimetry_radarConf2025}, we propose a framework for 3D imaging from the UAV nadir, paving the way toward utilizing lightweight compact radars on drones. This enables applications including vegetation detection, digital elevation mapping (DEM), search and rescue operations, and archaeology, among many others.

Specifically, in this paper, we present (to the best of our knowledge) the first InSAR framework for 3D radar imaging from the area directly below a UAV using a low-cost mm-wave radar. We extend the framework to account for the UAV attitude angles and propose an effective yet efficient method of phase error compensation across the virtual receive antennas of a multiple-input-multiple-output (MIMO) radar. Finally, the effectiveness of the proposed framework is evaluated through both simulation and experiment.

The remainder of this manuscript is organized as follows. Sec. \ref{sec:background}  presents the system model of imaging with an FMCW radar. Our proposed InSAR framework for generating 3D point clouds from a UAV nadir is described in Sec. \ref{sec:insar} with evaluation results presented in Sec. \ref{sec:results}. Finally, the paper is concluded in Sec. \ref{sec:Conclusion} along with future directions.	

\section{System Model}\label{sec:background}	
%The ISAR pipeline is supposed to produce a high-resolution (often) 2-D image of the moving target.
\subsection{FMCW radars}\label{sec:fmcw}
In FMCW radars, each Coherent Processing Interval (CPI) consists of transmitting $N_c$ chirps with pulse repetition interval (PRI) $T_c$. Each chirp is a continuous wave whose frequency starts at $f_c$ and increases linearly to $f_c+B$ where $B$ denotes the radar bandwidth. The transmitted chirp is modeled by:
\begin{equation}
	s_T(t) = a_c \exp\left[j\left(\omega_c+\frac{\gamma}{2}t\right)t\right],\, 0\le t \le T_c,
\end{equation}
where $a_c$ denotes the chirp amplitude, $\omega_c\triangleq 2\pi f_c$ is its starting angular frequency, and $\gamma$ is its slope.

The received echo signal from a specific scatterer $\bm{s}$ with round-trip time $\tau_{\bm{s}}$ is given by:
\begin{equation}
	s_R(t)=\sigma(\bm{s}) \exp\left[j\left(\omega_c+\frac{\gamma}{2}\left(t-\tau_{\bm{s}}\right)\right)\left(t-\tau_{\bm{s}}\right)\right],
\end{equation}
for $0\le t-\tau_{\bm{s}} \le T_c$, wherein the scatterer reflectivity, the system gains, and the propagation effects are included in $\sigma(\bm{s})$ for convenience. The received signal is demodulated to give the beat signal as follows:
\begin{equation}
	\begin{aligned}
		s(t)&=s_T^*(t)s_R(t)\\
		&=\sigma(\bm{s})\exp\left[-j\tau_{\bm{s}}\left(\omega_c+\gamma t-\frac{\gamma}{2}\tau_{\bm{s}}\right)\right],
	\end{aligned}		
\end{equation}
where $s_T^*(t)$ is the complex conjugate of $s_T(t)$.
The maximum unambiguous range of an FMCW radar (with I/Q receivers) is given by $r_{max}=\frac{\pi cF_s}{\gamma}$ where $F_s$ and $c$ denote the sampling rate and speed of light, respectively. This leads to $\frac{\gamma}{2}\tau_{\bm{s}}=\frac{\pi c F_s}{r_{max}}\times\frac{2r_s (t)}{c}=\frac{ r_s (t)}{r_{max}} 2\pi F_s$ where $r_s (t)$ is the slant range to the scatterer $\bm{s}$. This implies that the term $\gamma\tau_{\bm{s}}/2$ is negligible compared to $\omega_c$, especially since $F_s$ is at most several MHz\footnote{This term results in the residual video phase error (RVPE) which is negligible in the current technology of mm-wave FMCW radars.}.

Therefore, by collecting the beat signal of each received chirp in both fast-time and slow-time domains, the signal is approximated by:
\begin{equation}\label{eq:beat.signal}
	s(i,n)\approx\sigma(\bm{s})\exp\left[-j\frac{2}{c}\left(\omega_c+\gamma T_s i\right) r_s(n)\right],
\end{equation}
where $T_s\triangleq \frac{1}{F_s}$ denotes the receiver sampling period and the model has been discretized in fast time index $i\in\{0,1,\dots, T_c F_s-1\}$ and slow time index (viz. chirp number) $n\in\{0,1,\dots, N_c-1\}$.

\subsection{SAR imaging with FMCW radar}\label{sec:sar}
Diverse SAR algorithms offer different compromises between image quality and computational complexity. While the RD algorithm is limiting from the angular coverage aspect and BPA is computationally intensive, PFA strikes a balance by enhancing image quality through wavenumber distortion corrections while maintaining manageable complexity \cite{Doerry_PFA,Javadi_leca_2025}.
\begin{figure}[t]
	\begin{centering}
		\includegraphics[width=8cm]{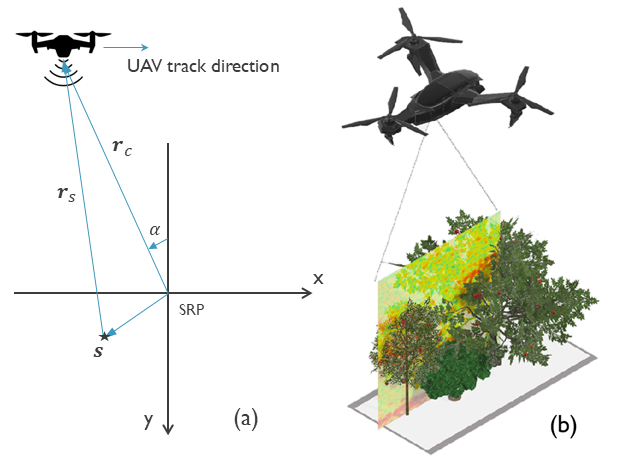}
		\par\end{centering}
	\caption{(a) The SAR 2D geometry. SRP indicates the SAR origin and stands for the SAR reference point. (b) The nadir plane of the UAV. \label{fig:geo}}
\end{figure}

Fig. \ref{fig:geo} shows the PFA geometry. Here, the radar is mounted below the UAV looking downward (along the $y$-axis) while the UAV flies along the $x$-axis. The slant range to a scatterer, $r_s(n)$, in \eqref{eq:beat.signal},  can be approximated by projecting the scatterer’s position onto the radar line of sight (RLOS) to the SAR reference point (SRP). This projection is expressed as:
\begin{equation}\label{eq:rs}
	r_s(n) = r_c(n)+y_s\cos\left(\alpha\right)+x_s\sin\left(\alpha\right),
\end{equation}
with $r_c(n)$ and $\alpha$ being the radar range to the SRP and its squint angle, respectively. Then, compensating the beat signal with respect to (w.r.t.) $r_c(n)$ gives
\begin{equation}\label{eq:compensated.signal}
	\begin{aligned}
		&s_C(i,n) \triangleq s(i,n)\exp\left[jk(i) r_c(n)\right]=\\
		&\int_{x_s} \int_{y_s} \sigma(x_s,y_s)\exp\left[-j\left(k_x(i,n)x_s+k_y(i,n)y_s\right)\right] dx_s dy_s,
	\end{aligned}
\end{equation}
wherein the integration is taken over all scatterers in the radar FoV. In \eqref{eq:compensated.signal}, $k(i) \triangleq \frac{2}{c}\left(\omega_c+\gamma T_s i\right)$ is the wavenumber (i.e., the spatial phase rate \cite{Doerry_PFA}), with components 
\[
k_x(i,n) \triangleq k(i) \sin\left(\alpha_n\right), \qquad
k_y(i,n) \triangleq k(i) \cos\left(\alpha_n\right),
\] 
where the dependence of the squint angle $\alpha$ on the chirp number $n$ is emphasized.

The signal model \eqref{eq:compensated.signal} has the form of a 2D Fourier transform (FT) but with co-dependent spatial frequencies (wavenumbers). Specifically, $k_x(i,n)$ and $k_y(i,n)$ form a polar format, where the amplitude is given by $k(i)$, and the phase varies with $n$. To enable fast Fourier transform (FFT), indices $i$ and $n$ are redefined into $i^\prime$ and $n^\prime$, respectively. This allows the range-compensated signal $s_C(i,n)$ to be interpolated onto a grid of wavenumbers, $k_x[n^\prime]$ and $k_y[i^\prime]$\footnote{Further details can be found in \cite{Doerry_PFA}.}. Consequently, an image of the target scene is reconstructed by performing a 2D inverse FFT (IFT) from the interpolated signal $s_C\left(i^\prime,n^\prime\right)$.
\begin{equation}
	\sigma(x_s,y_s)=2\text{D-IFT}\left\{s_C\left(i',n'\right)\right\}.
\end{equation}

\section{InSAR from the UAV nadir region}\label{sec:insar}
\subsection{Configuration}\label{sub:insar.fundamental}
We adopt the SAR geometry of Fig.~\ref{fig:geo}(a) for imaging in the UAV nadir plane, as depicted in Fig.~\ref{fig:geo}(b). In this geometry, the down-range lies in the imaging plane. This is equivalent to a zero grazing angle (or, equivalently, an incidence angle of $90^\circ$), which minimizes layover and foreshortening \cite{Doerry_notes}. It also
facilitates SAR imaging from the UAV nadir \cite{Javadi_depth_map_eurad2025}. Placing two receive antennas in the across-track dimension (i.e., along the $z$-axis)\footnote{In conventional InSAR configurations, the along-$z$ dimension is typically referred to as \emph{elevation}. However, in this paper, the $z$-axis does not represent the elevation of scatterers. Therefore, we refer to this dimension as \emph{across-track}, but we note that they are equivalent in InSAR terminology.} enables 3D imaging based on InSAR.

\begin{figure}[t]
	\begin{centering}
		\includegraphics[width=7cm]{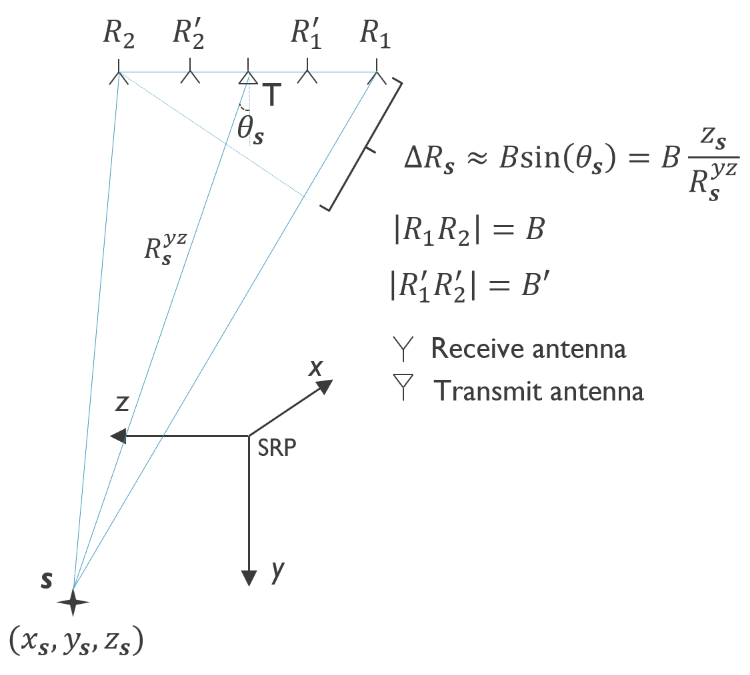}
		\par\end{centering}
	\caption{The InSAR geometry. \label{fig:insar.geometry}}
\end{figure}

Specifically, we place two receive antennas $R_1$ and $R_2$ with baseline distance $B$ and a transmitter $T$ (or equivalently, the zero-phase point) located at their midpoint. The baseline should be in the across-track direction, as shown in Fig.~\ref{fig:insar.geometry}. While large baselines improve robustness in across-track estimations \cite{Richards_InSAR_2007}, a small baseline is preferred to avoid the decorrelation of the phases at receive antennas \cite{Bamlery_1998}. We assume that the target scene is in the far-field of the antenna array so that wavefronts can be considered planar.

A second pair of receive antennas $R_1^{\prime}$ and $R_2^{\prime}$ with a different baseline $B'$ is leveraged in phase unwrapping, as explained in Sec.~\ref{sub:phase.unwrapping}. The two baselines are chosen so that their ratio is a rational number with coprime integers. We consider $B=b\frac{\lambda_n}{2}$ and $B'=b'\frac{\lambda_n}{2}$, where $\lambda_n$ is the radar nominal wavelength, and $b$ and $b'$ are coprime integers.

Assuming that the baseline $B$ is negligible compared to the slant range to the SRP,  it is straightforward to show that the beat signals of $R_1$ and $R_2$ corresponding to a scatterer $\bm{s}=(x_\subb{s},y_\subb{s},z_\subb{s})$ are related by
\begin{equation}\label{eq:interferometry.angle}
	s^{(R_1)}(i,n) = s^{(R_2)}(i,n) \exp\left(-j\frac{2\pi B}{\lambda}\sin \theta_\subb{s}\right),
\end{equation}
where $\lambda \triangleq \frac{c}{f_c}$ is the wavelength of the starting frequency and $\theta_\subb{s}$ denotes the elevation angle of the scatterer $\bm{s}$. Using the geometry of Fig.~\ref{fig:insar.geometry}, we have
\begin{equation}\label{eq:interferometry}
	s^{(R_1)}(i,n) = s^{(R_2)}(i,n) \exp\left(-j\frac{2\pi B}{\lambda} \frac{z_\subb{s}}{R_\subb{s}^{yz}}\right),
\end{equation}
wherein $R_\subb{s}^{yz} \triangleq \sqrt{z_\subb{s}^2+\left(y_\subb{s}-y_T\right)^2}$ with $y_T$ denoting the y coordinate of the radar transmitter (or, equivalently, the zero-phase point) on the UAV. Since the SRP is located beneath the zero-phase point, its $z$-coordinate is zero, i.e., $z_T=0$.  Given the phase difference of the scatterer $\bm{s}$ as $\psi \triangleq \frac{2\pi B}{\lambda} \frac{z_\subb{s}}{R_\subb{s}^{ys}}$ and defining $\kappa \triangleq \frac{\lambda \psi}{2 \pi B}$, the $z$-coordinate of $\bm{s}$ is given by
\begin{equation}\label{eq:z_s}
	z_\subb{s} = \pm\frac{R_y}{\sqrt{\frac{1}{\kappa^2}-1}},
\end{equation}
wherein $R_y \triangleq y_\subb{s} - y_T$ under the condition $\kappa^2<1$.

\begin{figure*}[ht]
	\begin{centering}
		\includegraphics[width=18cm]{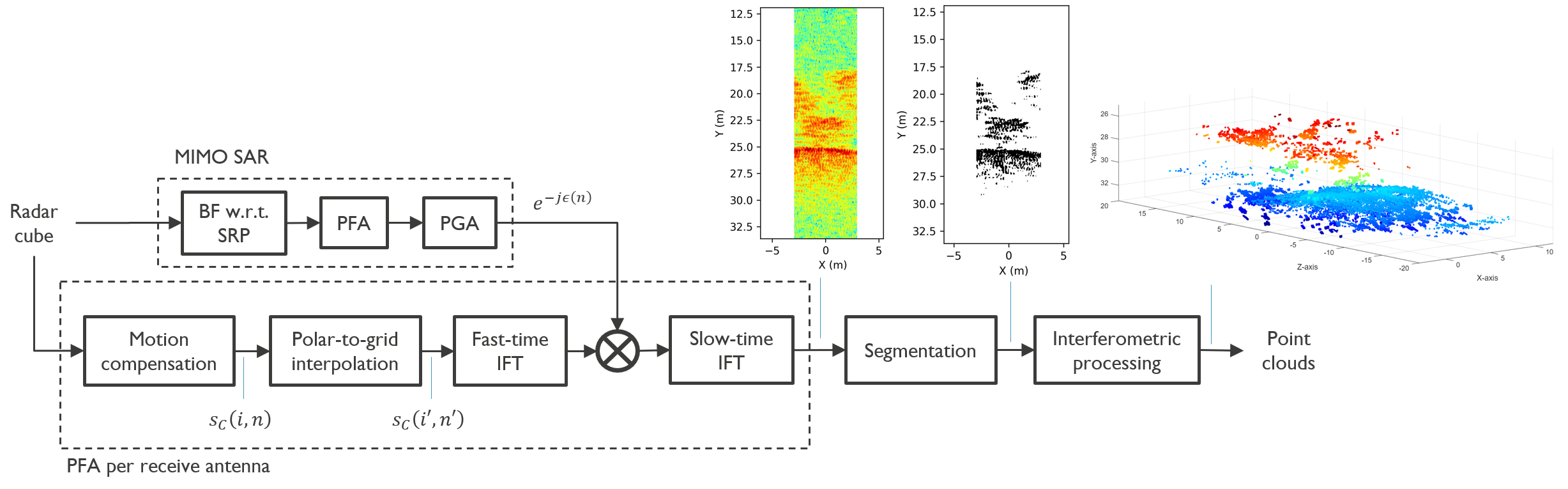}
		\par\end{centering}
	\caption{The InSAR framework for 3D imaging from the UAV nadir. $\epsilon(n)$ denotes the phase error. \label{fig:pipeline}}
\end{figure*}

While the $y$-coordinates of scatterers are obtained from the SAR image, Eq. \eqref{eq:z_s} calculates their $z$-coordinates. The following subsection outlines the essential steps for 3D radar imaging from the UAV nadir region.

\subsection{Pipeline}
Our proposed pipeline of 3D radar imaging is presented in Fig. ~\ref{fig:pipeline}. Each step is described below.

\subsubsection{Phase error compensation}
In airborne SAR, phase errors are primarily caused by positioning inaccuracies \cite{Doerry_BPA,Wahl1996} and significantly affect the accuracy of across-track estimation \cite{Richards_InSAR_2007,Bamlery_1998}. We denote the measured radar position by $[x_r, y_r, z_r]^T$ and assume it is subject to a positioning error $\epsilon$ along the along-track dimension $x$\footnote{Only the along-track positioning error is considered since the UAV has negligible motions in other directions during the SAR CPI.}. The true slant range to a scatterer located at $[x_s, y_s, 0]^T$ is given by:		
\begin{equation}\label{eq:rs.approx}
	\begin{aligned}
		r_s^\star &\triangleq \sqrt{\left(x_r+\epsilon-x_s\right)^2+(y_r-y_s)^2+z_r^2}\\
		&\approx r_s + \frac{x_r-x_s}{r_s}\epsilon\,
	\end{aligned}
\end{equation}
where $r_s \triangleq \sqrt{\left(x_r-x_s\right)^2+(y_r-y_s)^2+z_r^2}$ represents the slant range computed using the erroneous radar position. For notational simplicity, the dependence of both the position and the associated error on the slow-time index $n$ is omitted. Substituting \eqref{eq:rs} into  \eqref{eq:rs.approx} and using the approximation $\frac{x_r(n)-x_s}{r_s(n)}\approx\sin(\alpha_n)$ gives the compensated beat signal as \cite{Javadi_leca_2025}:
\begin{equation}\label{eq:motion.error.model}
	s_C^\star(i,n) = \exp\left[jk_x(i,n) \epsilon(n)\right]s_C(i,n),
\end{equation}
which shows that the positioning error $\epsilon(n)$ manifests as a phase error in the compensated beat signal with respect to the range to the SRP, when using the \emph{measured} platform positions. This result is consistent with the phase error model adopted in the phase gradient autofocus (PGA) algorithm \cite{Wahl1996}.

The PGA algorithm assumes that each range bin in the SAR image is dominated by at most one strong scatterer and demonstrates that the phase error spectrum is modulated by this dominant scatterer. It then estimates the phase error gradient by exploiting redundancy across multiple dominant scatterers within the image. The phase error itself is subsequently recovered by integrating the estimated gradient. We adopt PGA for phase compensation because it does not involve intensive optimization, making it a preferred algorithm for resource-constrained UAVs. However, PGA implicitly assumes that the phase error depends solely on the slow-time index. When the target scene is extensive or the SAR aperture is large enough to violate this assumption, PGA can be replaced with a more appropriate phase compensation algorithm that is better suited to the specific SAR imaging scenario \cite{mea2014,Xing2009}.

To mitigate phase errors in UAV-borne SAR, we adopt a MIMO-SAR processing strategy. First, beamforming (BF) across antenna elements is performed with respect to the SRP at the UAV nadir, using inertial measurement unit (IMU) data, as detailed in \cite{Javadi_depth_map_eurad2025}. BF coherently integrates the signals received by the antenna elements to improve signal-to-noise ratio (SNR), thereby enabling more accurate phase error estimation. Next, SAR imaging is performed using PFA. Finally, PGA estimates the residual phase error, which is then used to correct the phase of SAR images reconstructed at each receive antenna.

Accordingly, as shown in Fig. \ref{fig:pipeline}, the phase error for each receive antenna is compensated by multiplying the range profile by $\exp\left[-j\epsilon(n)\right]$ with $\epsilon(n)$.

\subsubsection{SAR per receive antenna}
After the phase error is compensated for in the range profiles of the receive antennas, SAR images are given by applying a second IFT in slow-time, as depicted in Fig.~\ref{fig:pipeline}. It is worth mentioning that other SAR algorithms may be adopted depending on the scenario. We resort to PFA since it provides an appropriate compromise between performance and complexity \cite{Javadi_leca_2025}. 

\subsubsection{Segmentation}
To determine dominant scatterers, we employ the Rayleigh-based Segmentation (RaySe) method introduced in \cite{Javadi2022}. RaySe leverages the fact that the background noise amplitude in a SAR image follows a Rayleigh-based distribution. It estimates the Rayleigh distribution parameter from the variance of the SAR image amplitude and sets a detection threshold in the upper tail of the distribution, thereby isolating dominant scatterers.

The SAR images of the receive antennas are thresholded using RaySe. The bins detected in all images are intersected, giving the $(x,y)$ coordinates of the major scatterers. The goal is to obtain their $z$-coordinates through interferometric processing. We denote the thresholded SAR images of the receive antennas $R_1$, $R_2$, $R_1^{\prime}$, and $R_2^{\prime}$ by $S_{R_1}$, $S_{R_2}$, $S_{R_1^{\prime}}$, and $S_{R_2^{\prime}}$, respectively.

\subsubsection{Interferometric processing}\label{sub:phase.unwrapping}
Interferometric processing consists of the following steps.
\begin{enumerate}
	\item \textbf{Registration of SAR images:} The across-track estimation depends on comparing phase differences from matching pixels in two apertures. To avoid any distortions due to misalignment of the SAR images, it is essential to ensure that the images are accurately registered.
	
	The across-track coordinate of a detected scatterer is derived from \eqref{eq:interferometry.angle} under the assumption that $\theta_\subb{s}$ remains invariant throughout the SAR CPI. This assumption, however, does not hold if the UAV experiences roll variations during the CPI, in which case $\theta_\subb{s}$ becomes time-varying and the interferometric phase becomes a function of the pulse index $n$. Such dependency can cause geometric distortions between the images, rendering pixel-wise phase differences unsuitable for accurate across-track estimation.
	
	Specifically, if the roll varies with rate $\theta'$, $\theta_\subb{s}$ in Eq. (7) can be replaced with $\theta_{\bm{s},0}+\theta' T_c n$, giving:
	\begin{equation}\label{eq:registration}
		\begin{aligned}
			s^{(R_1)}(i,n) &= s^{(R_2)}(i,n) \exp\left[-j\frac{2\pi B}{\lambda}\sin \left(\theta_{\bm{s},0}+\theta' T_c n\right)\right]\\
			&\approx s^{(R_2)}(i,n)\exp(j \zeta n) \exp\left(-j\frac{2\pi B}{\lambda}\sin \theta_{\bm{s},0}\right),			
		\end{aligned}		
	\end{equation}
	where $\zeta \triangleq \frac{2\pi B}{\lambda} \cos \left(\theta_{\bm{s},0}\right) \theta' T_c$ and $\theta_{\bm{s},0}$ is the initial value of $\theta_\subb{s}$. The above equation implies that the roll variation induces a shift of the SAR image of $R_2$ relative to $R_1$ along the cross-range dimension.
	
	To demonstrate the impact, assume $\alpha_n = \alpha' T_c n$ and multiply the compensated beat signal in \eqref{eq:compensated.signal} by $\exp(j \zeta n)$. For $i=0$ and a single scatterer, this gives
	\begin{equation}
		\begin{aligned}
			&s_C\left(0,n\right) = \sigma\left(x_s,y_s\right)\times\\
			&\exp\left\{-j\left[\frac{2\pi T_c}{\lambda}\left(2\alpha'x_s-B\cos\left(\theta_{\bm{s},0}\right)\theta'\right)n+\frac{4\pi}{\lambda}y_s\right]\right\}.
		\end{aligned}		
	\end{equation}
	
	Hence, the cross-range dimension is shifted by $\frac{B \cos\left(\theta_{\bm{s},0}\right)}{2\alpha'}\theta'$. If this shift is less than the cross-range resolution, misregistration is negligible, meaning that the two images are registered. At a UAV height of approximately $r_c$, the cross-range resolution is approximately $\frac{\lambda}{2\alpha' N_c T_c}$. Therefore, the condition for no misregistration is:
	\begin{equation}\label{eq.misregistration.criterion}
		\theta' < \frac{\lambda}{B \left(N_c T_c\right)}\,
	\end{equation}
	wherein $\cos\left(\theta_{\bm{s},0}\right)$ is replaced with its maximum value, $1$. The misregistration criterion \eqref{eq.misregistration.criterion} indicates that the roll variation rate limit is inversely proportional to both baseline and the SAR CPI. This means that for a small CPI and a short baseline, the SAR images are well registered with each other and hence, no registration step is required. However, if the roll variation rate exceeds this limit, the resulting misregistration must be compensated prior to the interferometric processing using appropriate methods, such as the cross-correlation-based approach in \cite{Richards_InSAR_2007} or the rotation-estimation-based method in \cite{InISAR_Zhang2004}. In the experimental setup used in this paper (Table~\ref{tab:radar.setting}), the limit is very high, implying no misregistration.
	\item \textbf{Interferometry:} The pixel-wise phase difference between each pair of images is calculated by
	\begin{equation}\label{eq:phases}
		\begin{split}
			\phi = \angle S_{R_1}\odot S_{R_2}^{*},\\
			\phi^{\prime} = \angle S_{R_1^{\prime}} \odot S_{R_2^{\prime}}^{*},
		\end{split}
	\end{equation}
	where $(\cdot)^{*}$ denotes complex conjugation and $\odot$ denotes element-wise multiplication.
	
	\item \textbf{Phase unwrapping:} The phases given by \eqref{eq:phases} are wrapped in $(-2\pi,2\pi]$ and need to be unwrapped to recover the actual phase differences. To this end, we resort to the multi-baseline phase unwrapping (MBPU) method in \cite{mbpu_lin_2022}, where two pairs of receive antennas with relatively prime baselines are used, as illustrated in Fig.~\ref{fig:insar.geometry}. From \eqref{eq:interferometry}, the unwrapped phases $\psi = \phi + 2\pi k$ and $\psi'=\phi'+2\pi k'$ ($k$ and $k'$ are integer ambiguity numbers) are related by
	\begin{equation}\label{eq:ambiguity.numbers.relation}
		k=\frac{B}{B'}k'+\frac{1}{2\pi}\left(\frac{B}{B'}\phi'-\phi\right).
	\end{equation}		
	Then, the ambiguity numbers are calculated by
	\begin{equation}
		k,k'=\arg \underset{k,k'\in\mathbb{Z}}{\min} \left|\frac{B}{B'}k'-k+\frac{1}{2\pi}\left(\frac{B}{B'}\phi'-\phi\right)\right|.
	\end{equation}
	
	The MBPU algorithm in \cite{mbpu_lin_2022} includes an extra clustering step of the phase differences of the pixels based on the intercept terms of \eqref{eq:ambiguity.numbers.relation} to reduce the impact of additive and phase noises. We omit the clustering step because of its considerable complexity. Furthermore, the noise impact is reduced significantly by the compensation procedure explained earlier.
	
	\item \textbf{Across-track estimation: } After the unambiguous values are estimated, the unwrapped phase difference is calculated using $\psi=\phi + 2\pi k$. Since the UAV moves along the $x$-axis, $y_T$ has limited variations; therefore, we use the average of its values in slow time, $\bar{y_T}$. Furthermore, during the SAR CPI, we have $z_T \approx 0$ since the target scene is located beneath the zero-phase point $T$. Then, the $z$-coordinate is estimated by \eqref{eq:z_s}. The positive sign is chosen when $\psi\ge0$; otherwise, the negative sign applies.
	
\end{enumerate}	

\subsubsection*{Accounting for UAV attitude}
To derive \eqref{eq:z_s}, we implicitly assumed that the UAV moves along the $x$-axis with no changes in attitude. In practice, the UAV attitude may vary during flight to maintain a pre-planned trajectory. To account for these attitude changes, we define the SAR coordinate system based on the UAV flight path and analyze the impact on the InSAR measurement.

Specifically, the UAV flies from the start point to the end point of the aperture with specific yaw and pitch angles. Over the duration of the CPI (e.g., a fraction of a second, such as about $27$ ms in our experiment), variations in yaw and pitch are negligible. The yaw defines the UAV heading, which determines the $x$-axis, while the pitch determines the $y$-axis. Consequently, the directions of $x$- and $y$-axes adapt with variations in yaw and pitch and hence, they do not affect the InSAR geometry. However, the UAV roll impacts the configuration in Fig. \ref{fig:insar.geometry}. With a roll $\rho$, the across-track equation evolves to:	
\begin{equation}\label{eq:z_s.with.roll}
	z_\subb{s} = R_y\frac{ \sin \rho \cos \rho \pm \sqrt{\kappa^2 \sin^2 \rho + \cos^2 \rho}}{\kappa^2 - \cos^2 \rho} \,.
\end{equation}

\section{Evaluation results}\label{sec:results}
In this section, we present the evaluation results of our proposed framework for 3D imaging from a UAV nadir in both simulated and experimental scenarios.

\subsection{Simulation results}
The simulation scenario uses a MIMO radar comprising two transmitters and four receivers, forming a linear virtual array of eight receive antennas with inter-antenna spacing $\frac{\lambda}{2}=2.5$~mm. We use the two receive antennas at the two ends of the virtual array as the primary pair (i.e., $R_1$ and $R_2$) for InSAR and another pair with spacing $5\frac{\lambda}{2}$ in the middle as the secondary pair (i.e., $R^{\prime}_1$ and $R^{\prime}_2$) for phase unwrapping. This is equivalent to the configuration depicted in the bottom linear array of Fig.~\ref{fig:setup}(c). Table~\ref{tab:radar.setting} lists the radar parameters. 

\begin{table}
	\centering
	\begin{threeparttable}
		\caption{Radar parameters used in both simulation and experimental scenarios.}
		
		\begin{tabular}{ccccccc}\label{tab:radar.setting}
			$f_c$ & Bandwidth & PRI\tnote{1} ($T_c$)  &  $N_c$ & $T_f$\tnote{2} & $S$\tnote{3} & $F_s$\tnote{4}\\
			(GHz) & (GHz) &  ($\mu$s) &  & (ms) & & (MHz)\\
			\hline
			\hline
			60 & 1.442    & 102.88    & 255 & 100  & 400    & 8  \\
			\hline
		\end{tabular}
		\begin{tablenotes}
			\item[1] Pulse repetition interval.
			\item [2] Radar frame period.
			\item[3] Number of samples per chirp.
			\item[4] Receiver sampling frequency.
		\end{tablenotes}
	\end{threeparttable}
\end{table}

The antenna array is oriented along the $z$-axis, points downward, and flies at a constant velocity of $7$~m/s at an altitude of $25$~m above ground level. Several scatterers are randomly deployed within the area below the radar.  

The positioning error due to an along-track velocity perturbation $v_e$ is given by $\epsilon(n) = v_e T_c n$ resulting in the phase error $k_x(i,n) \epsilon(n) = k(i) \sin(\alpha_n) v_e T_c n$. By neglecting the variation of $k(i)$ over a PRI and approximating $\sin(\alpha_n)\approx \alpha'n$, the phase error becomes quadratic in $n$, i.e., $k_x(i,n) \epsilon(n) \approx k \alpha' v_e T_c n^2$. This constitutes the dominant phase error in SAR imaging \cite{ghiglia1998two,Wahl1996} and means that inaccuracies in velocity measurements are the primary source of phase errors in SAR imagery \cite{Manzoni2023,Javadi_leca_2025}.

\begin{figure}[ht]
	\begin{centering}
		\includegraphics[width=5cm]{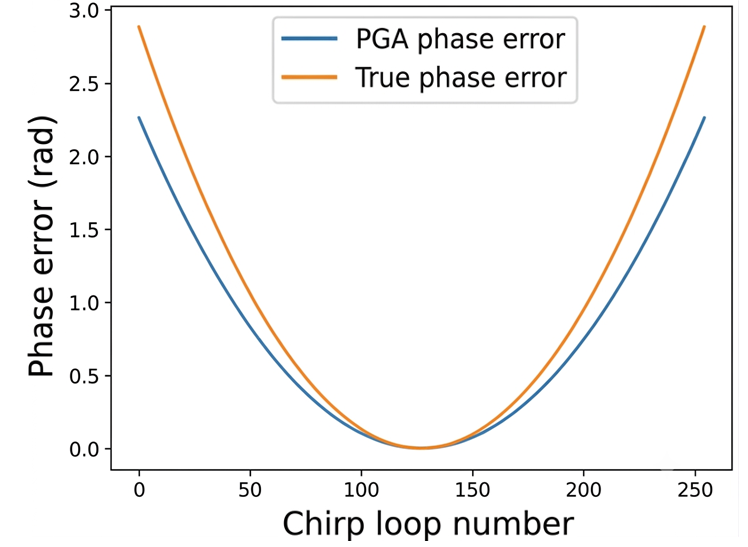}
		\par\end{centering}
	\caption{The quadratic phase error and its estimation by PGA. \label{fig:pga}}
\end{figure}

Accordingly, to emulate a condition with significant phase distortions, we introduce a velocity perturbation of $2.5$~m/s along the $x$-axis and $0.1$~m/s along the remaining axes\footnote{Commercial UAV IMU/GPS systems typically achieve a velocity accuracy below $0.5$~m/s (e.g., \cite{ublox_M8_datasheet}).}. Fig. \ref{fig:pga} depicts the simulated phase error and the quadratic part of the phase error estimated by PGA.	As shown, PGA significantly compensates for the phase error. The slight discrepancy between the phase error and the PGA estimate is primarily due to the large spatial extent of the scatterers along the $z$-axis compared to the UAV altitude, as well as the extended SAR aperture caused by the relatively large velocity error. These factors render the phase error range-variant \cite{mea2014,Xing2009}, which deviates from the PGA assumption and leads to imperfect phase error compensation. This limitation can be mitigated by employing algorithms such as minimum entropy autofocus (MEA) \cite{mea2014} or geometry-based methods \cite{Xing2009,Grassi2026}, albeit at the cost of increased computational complexity.

\begin{figure}[ht]
	\begin{centering}
		\includegraphics[width=\columnwidth]{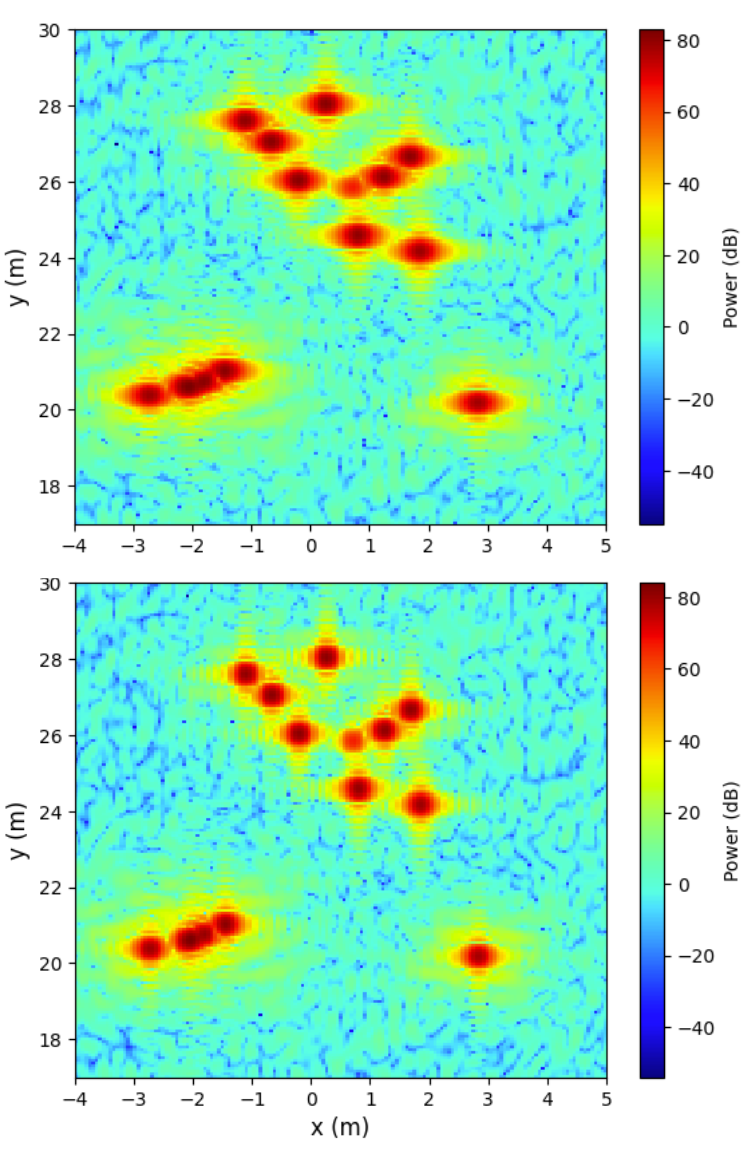}
		\par\end{centering}
	\caption{SAR image reconstructed at receiver $R_1$ without (top) and with (bottom) phase-error compensation. \label{fig:simulation.af}}
\end{figure}

The impact of phase-error compensation on SAR imaging at $R_1$ is illustrated in Fig.~\ref{fig:simulation.af}. As shown, the SAR image becomes better focused.

\begin{figure}[ht]
	\begin{centering}
		\includegraphics[width=9cm]{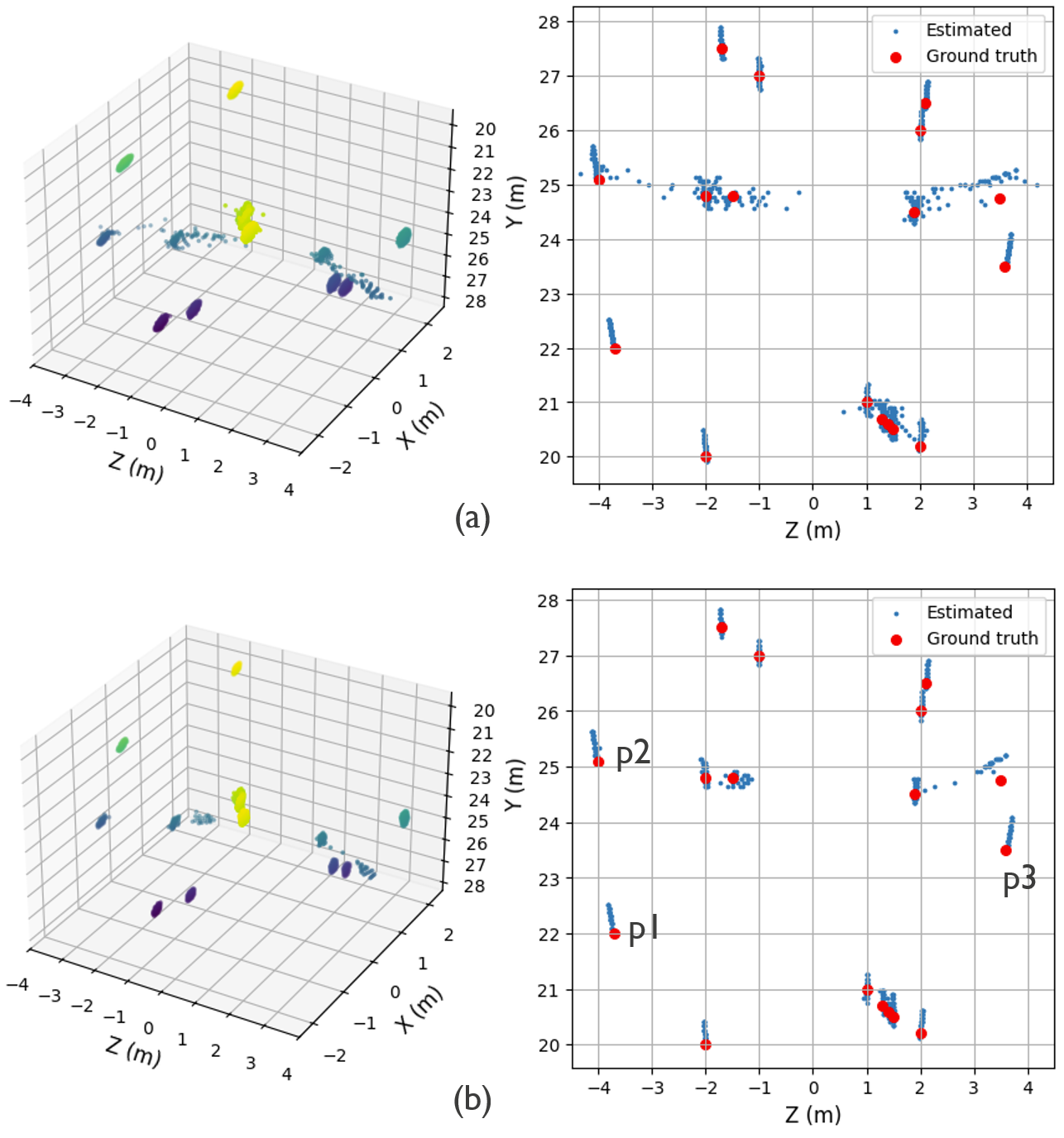}
		\par\end{centering}
	\caption{Simulation results of 3D imaging. Left: Point clouds given by the InSAR pipeline. Right: Point clouds projected onto the $y$--$z$ plane. (a) Without phase-error compensation. (b) With phase-error compensation, as in the pipeline of Fig.~\ref{fig:pipeline}. \label{fig:simulation.result}}
\end{figure}

Fig.~\ref{fig:simulation.result} presents the simulation outcomes for two scenarios with and without phase-error compensation. Here, the $y$-axis has been reversed to depict range from the radar, which was flying $25$~m above the ground surface. As shown, the proposed pipeline in Fig.~\ref{fig:pipeline} effectively mitigates phase errors, resulting in more focused point clouds per point scatterer. Notably, the compensation improves not only the range-azimuth ($x$–$y$) resolution but also enhances resolution along the across-track dimension.

\begin{figure}[ht]
	\begin{centering}
		\includegraphics[width=7cm]{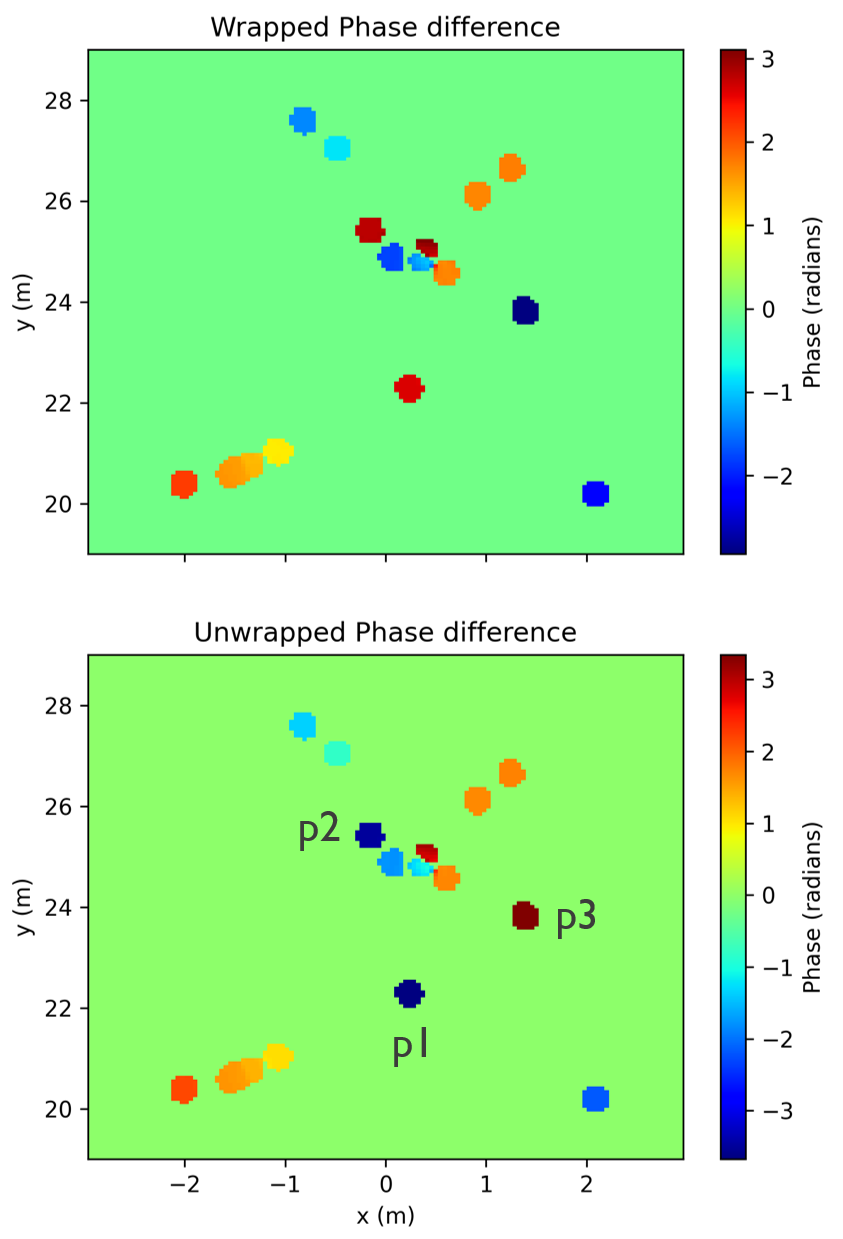}
		\par\end{centering}
	\caption{Pixel-wise phase difference between SAR images of $R_1$ and $R_2$ before (top) and after (bottom) phase unwrapping. The marked points indicate corrected phases for higher $z$ values, consistent with Fig.~\ref{fig:simulation.result}.}  \label{fig:simulation.unwrap}
\end{figure}

Additionally, Fig.~\ref{fig:simulation.unwrap} highlights the significance of phase unwrapping, particularly for correctly estimating larger $z$ values. Three points are marked in the bottom plots of Fig.~\ref{fig:simulation.result} and \ref{fig:simulation.unwrap}. Their phase differences are wrapped since their values lie outside the range $(-2\pi,2\pi]$.

The accuracy of the proposed 3D SAR processing pipeline is assessed by imaging scenes with varying numbers of point scatterers randomly distributed within a cubic region of size $10$ m centered around the SRP. For each scatterer count, the experiment is repeated for $10$ independent trials. The root mean square error (RMSE) is computed across both the $10$ iterations and all scatterers, based on the minimum Euclidean distance between each ground-truth scatterer location and its closest point in the reconstructed point cloud.

\begin{figure}[ht]
	\begin{centering}
		\includegraphics[width=8cm]{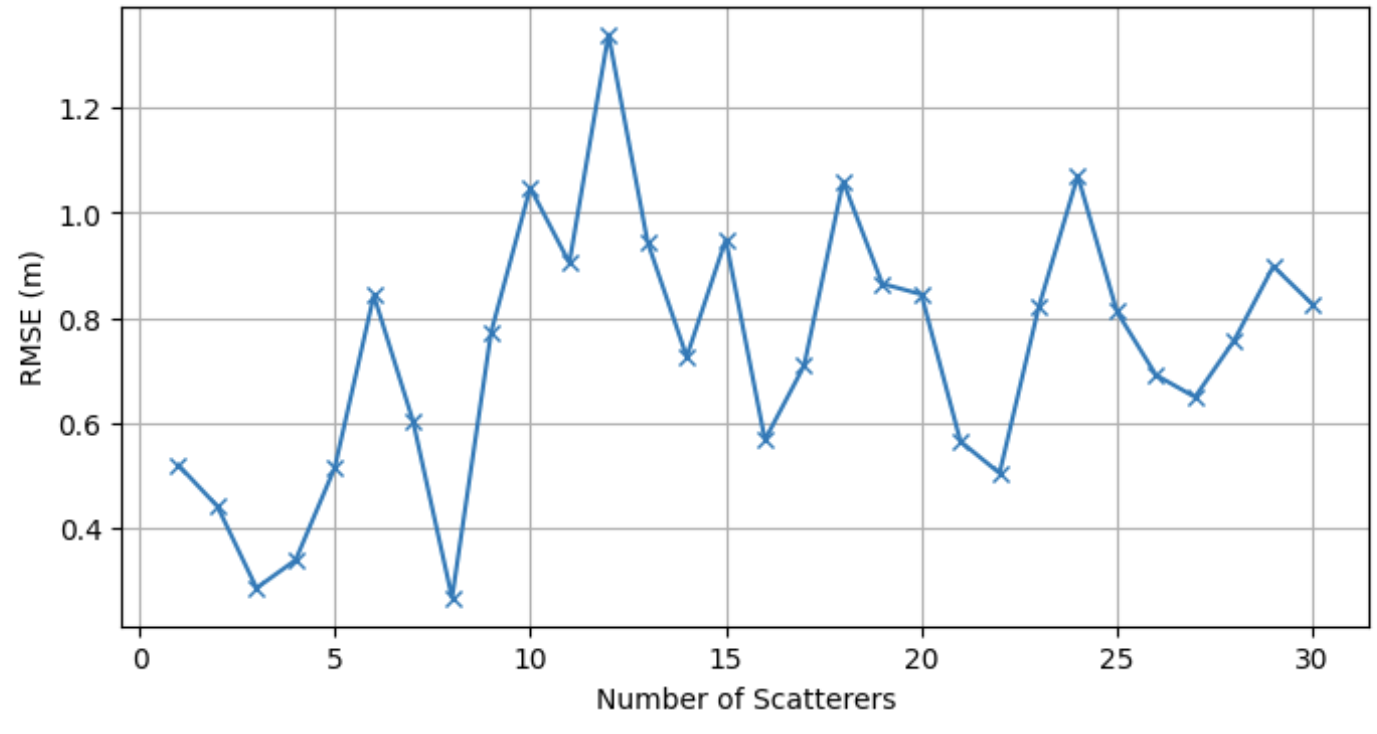}
		\par\end{centering}
	\caption{RMSE of the proposed 3D imaging framework for different number of scatterers randomly deployed around the SRP.  \label{fig:rmse}}
\end{figure}

In each trial, a random along-track velocity error uniformly distributed between $-2$ m/s and $2$ m/s, as well as a random roll perturbation between $-5^\circ$ and $5^\circ$, is injected to evaluate robustness against motion and attitude uncertainties. Fig.~\ref{fig:rmse} presents the resulting RMSE performance for different target scene configurations. As the number of scatterers increases, the RMSE remains consistently bounded within a sub-meter range. The stable and bounded RMSE across diverse scene configurations confirms the robustness of the proposed 3D SAR imaging pipeline under motion and attitude perturbations.

It is worth mentioning that the simulation results provided here are primarily intended to validate the proposed 3D InSAR pipeline under representative conditions, rather than to exhaustively explore all system parameters. In particular, our current simulations do not separately quantify the effects of varying factors such as SNR, the magnitude of residual phase errors, reduced range or cross-range resolution, or potential phase-unwrapping errors. These parameters can influence scatterer detection, image focusing, interferometric phase estimation, and ultimately the accuracy of the reconstructed point cloud in practice. For example, lower SNR or severe residual phase errors could degrade SAR image focus and interferometric measurement accuracy, while significantly reduced resolution might increase scatterer mis-association or bias the across-track estimates. Likewise, unaddressed phase unwrapping errors may produce large outliers in the across-track domain. A full parametric sensitivity analysis of these factors is beyond the scope of the present work, but it represents an important avenue for future research.

\subsection{Experimental results}
\begin{figure}[t]
	\begin{centering}
		\includegraphics[width=8.5cm]{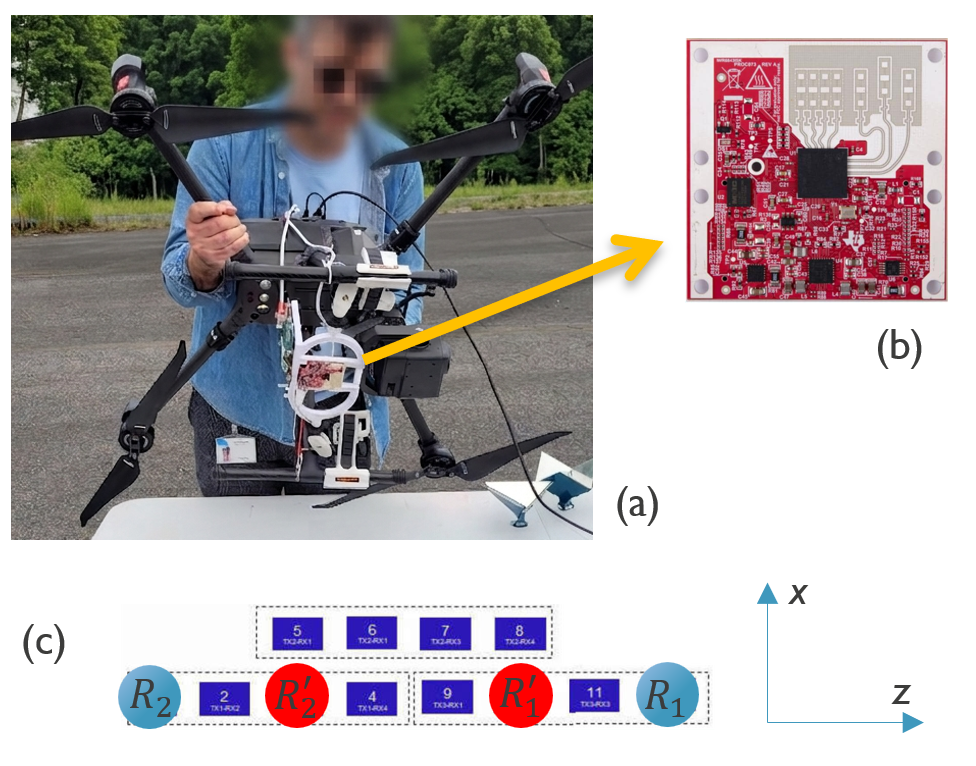}
		\par\end{centering}
	\caption{Experiment setup. (a) Radar mounted beneath the UAV. (b) MIMO FMCW radar board. (c) Virtual receive array and its placement with regard to the $x$- and $z$-axes. The two pairs of antennas used for InSAR processing are highlighted in red and light blue.  \label{fig:setup}}
\end{figure}

In this section, we present a qualitative evaluation of the proposed pipeline in practical scenarios. To that end, we use data from a MIMO FMCW radar (IWR6843ISK of Texas Instruments) installed on the underside of a drone (DJI Matrice-300), pointing downward, with the configuration shown in Fig.~\ref{fig:setup}. The radar features three transmit (Tx) and four receive (Rx) antennas, forming 12 virtual channels arranged in two rows aligned with the cross-track direction. It operates in the 60~GHz band, with the parameters summarized in Table~\ref{tab:radar.setting}.
In addition to the radar, the UAV carried a Zenmuse H20 camera to provide optical imagery for visual validation. The UAV flew at a nominal speed of $7$~m/s and an altitude of approximately $25$~m during data collection.

\begin{figure*}[t]
	\begin{centering}
		\includegraphics[width=17cm]{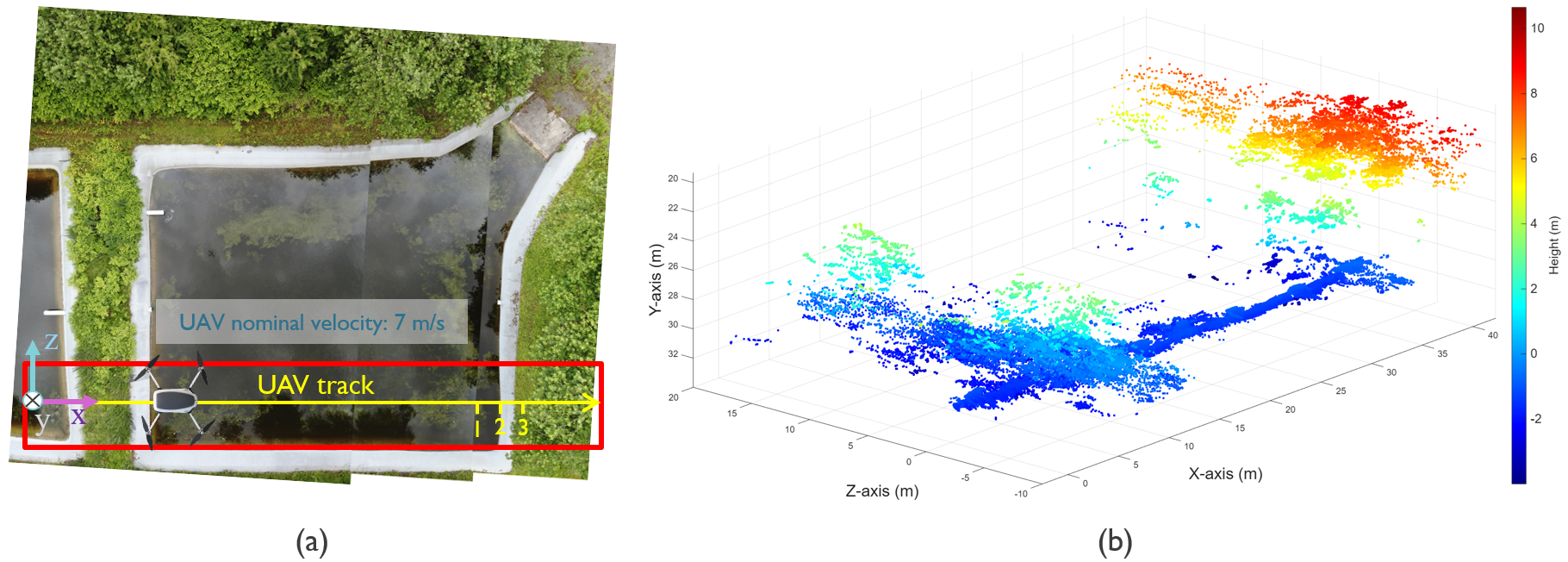}
		\par\end{centering}
	\caption{Example of 3D radar imaging from above a pond. (a) UAV flight over the pond surrounded by trees and vegetation. The x-axis is along the flight direction while the y-axis points downward. Instances 1, 2, and 3 are specified for demonstrating the impact of phase error compensation in Fig. \ref{fig:2dsar}. (b) Point clouds given by the proposed 3D InSAR framework.  \label{fig:experiment}}
\end{figure*}

\begin{figure*}[ht]
	\begin{centering}
		\includegraphics[width=12cm]{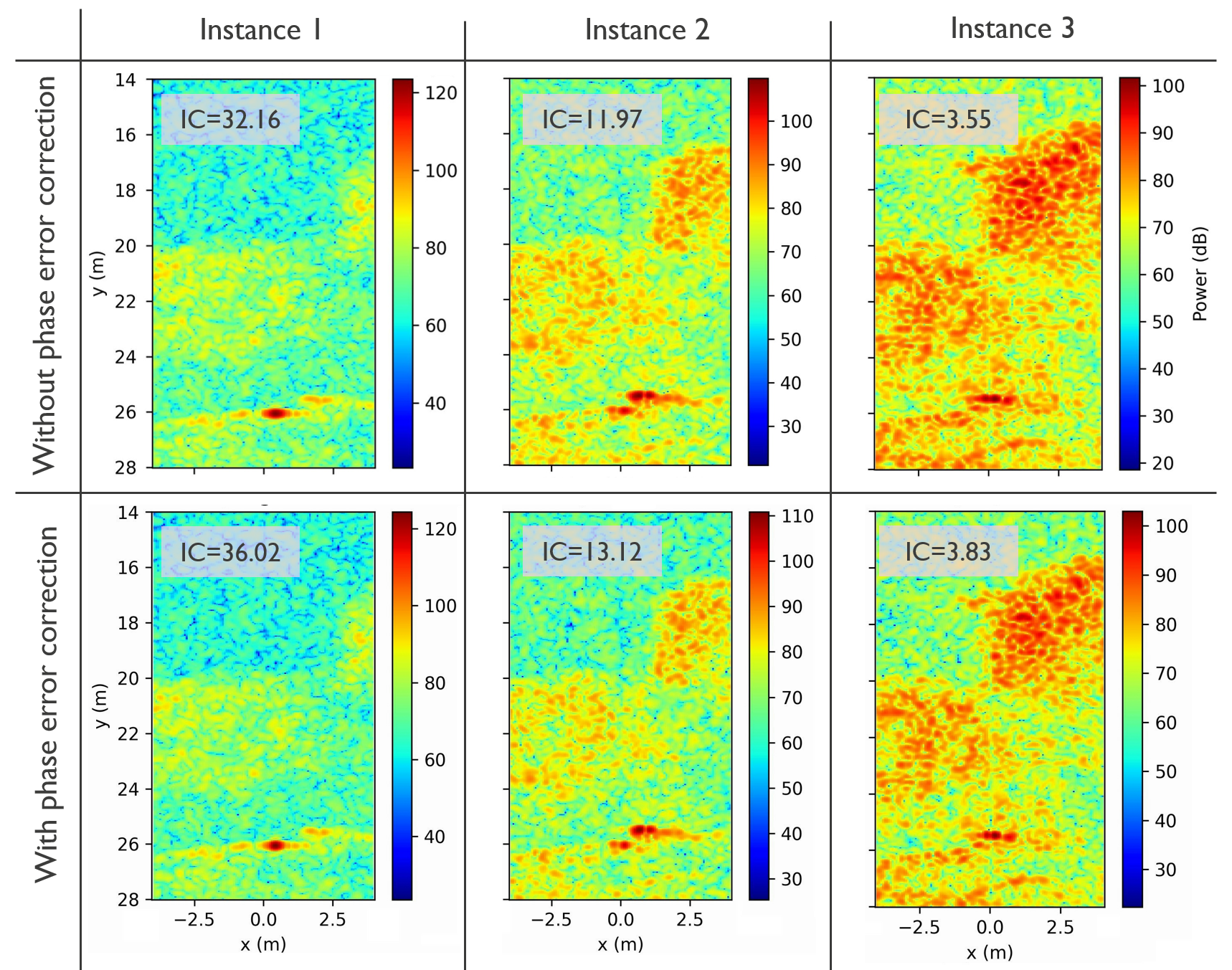}
		\par\end{centering}
	\caption{The impact of phase error compensation on the SAR images of a channel at three instances specified in Fig. \ref{fig:experiment}. The images qualities have been quantified in terms of image contrast (IC).  \label{fig:2dsar}}
\end{figure*}

\begin{figure}[ht]
	\begin{centering}
		\includegraphics[width=8.5cm]{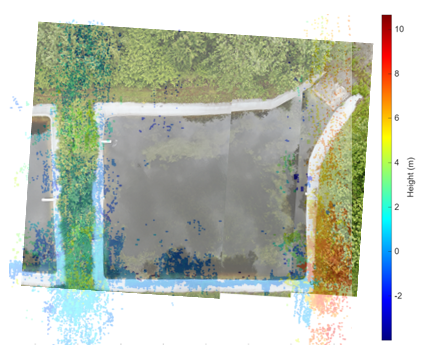}
		\par\end{centering}
	\caption{Point clouds of Fig. \ref{fig:experiment}(b) overlaid on the pond image. \label{fig:experiment.overlaid}}
\end{figure}

The experimental scenario, as shown in the bird's-eye view (BEV) image in Fig. \ref{fig:experiment}(a), involves flying the UAV over a pond divided into two sections. The UAV heading defines the x-axis, the y-axis points downward, and the z-axis corresponds to the across-track direction. The UAV flies at a nominal velocity of $7$~m/s and is equipped with GPS that reported the UAV position at a rate of $10$~Hz. This update rate is significantly smaller than the radar pulse repetition frequency (about $3.42$~KHz). This necessitates the interpolation of the UAV position for each chirp, which introduces phase errors due to positioning inaccuracies.

Fig. \ref{fig:2dsar} depicts the 2D SAR images reconstructed at a selected virtual antenna at three time instances, with and without compensating the phase errors. The three instants, indicated in Fig. \ref{fig:experiment}(a) correspond to moments when the UAV approaches the pond end edge. The SAR images were reconstructed along the depth dimension and capture the progressive scene evolution: the emergence of the pond edge at instance 1, followed by the gradual appearance of tree foliage at instances 2 and 3. As can be seen, phase error compensation improves the image focus, as confirmed by the enhancement of the image contrast (IC) defined as the normalized image variance \cite{Martorella2005}:

\begin{equation}\label{eq:ic}
	IC\left(\sigma(x,y)\right)\triangleq\frac{\sqrt{\text{mean}\left\{\left(I-\text{mean}(I)\right)^2\right\}}}{\text{mean}(I)},
\end{equation}
with $I\triangleq\left|\sigma(x,y)\right|^2$ being the image intensity.

Considering the interferometric setup shown in Fig. \ref{fig:experiment}(b) and the radar parameters in Table \ref{tab:radar.setting}, the misregistration threshold is approximately $624$ deg/s, whereas the maximum roll variation rate of the UAV during its flight is $8$ deg/s. Therefore, according to criterion \eqref{eq.misregistration.criterion}, no image registration step is required.

Interferometric processing at the end of each CPI yields point clouds in the coordinate system corresponding to that CPI. We aggregate the point clouds of all CPIs by transforming them into the coordinate system of the first CPI. Fig. \ref{fig:experiment}(b) illustrates the point clouds from the UAV nadir, captured and aggregated over the flight duration. In the left part of Fig. \ref{fig:experiment}(b), the point clouds indicate two trees (in cyan tones) at both ends of the divider. Additionally, several tall trees are visible on the right side of the pond. The edges of the pond are also well reconstructed. Note that the water surface behaves like a specular reflector, scattering the radar signals away rather than reflecting them back to the radar. Therefore, it is barely detected by the radar. 

Finally, Fig.~\ref{fig:experiment.overlaid} shows the point clouds overlaid on the BEV image. To this end, radar–camera extrinsic calibration was performed by placing multiple corner reflectors on large checkerboards at different ground locations. The reflectors were detected by the radar, and the checkerboards were observed by the camera. The calibration transformation matrix was then estimated using at least four corresponding radar–camera detection pairs.

A video corresponding to the experimental result is available in the Multimedia Material associated with this paper.

\section{Conclusion}\label{sec:Conclusion}
In this paper, we proposed a framework for capturing 3D point clouds from the region directly beneath a UAV using a MIMO FMCW radar. Unlike common UAV SAR configurations, we placed the antenna array on the underside of the UAV oriented downward, with the antenna array aligned along the across-track dimension. SAR imaging is then performed in the UAV nadir plane. Consequently, the 3D point clouds are obtained by interferometric processing across two pairs of receive antennas. Additionally, our proposed pipeline incorporates a method for compensating radar phase errors, whose effectiveness was confirmed through simulation. The framework was evaluated in practice by conducting a flight over a pond where it successfully reconstructed the pond boundaries as well as the surrounding trees and vegetation. The point clouds can serve as inputs to rescue operation models, support 3D scene reconstruction applications, and enable high-resolution multimodal fusion for object detection purposes.

\section{Acknowledgement}
The research leading to these results received funding from the Horizon Europe project \emph{Edge AI Technologies for Optimised Performance Embedded Processing} (Grant agreement ID: 101097300).

\bibliography{keylatex}

@misc{ublox_M8_datasheet,
	author = {{u-blox AG}},
	title = {{NEO-M8} u-blox {M8} concurrent {GNSS} modules},
	year = {2025},
	howpublished = {\url{https://content.u-blox.com/sites/default/files/NEO-M8-FW3_DataSheet_UBX-15031086.pdf}}
}

@ARTICLE{Ausherman1984,
	author={Ausherman, Dale A. and Kozma, Adam and Walker, Jack L. and Jones, Harrison M. and Poggio, Enrico C.},
	journal={IEEE Transactions on Aerospace and Electronic Systems}, 
	title={Developments in Radar Imaging}, 
	year={1984},
	volume={AES-20},
	number={4},
	pages={363-400},
	doi={10.1109/TAES.1984.4502060}}

@ARTICLE{Martorella2005,
	author = {M. Martorella and F. Berizzi and B. Haywood},
	title = {Contrast maximisation based technique for {2-D ISAR} autofocusing},
	journal = {IEE Proceedings - Radar, Sonar and Navigation},
	issue = {4},   
	volume = {152},
	year = {2005},
	month = {August},
	pages = {253-262(9)},
}

@book{Franceschetti1999,
	author = {G. Franceschetti and R. Lanari},
	title = {Synthetic Aperture Radar Processing},
	publisher = {CRC Press},
	year = {1999},
	type = {Book}
}

@ARTICLE{Wahl1996,  author={Wahl, D.E. and Eichel, P.H. and Ghiglia, D.C. and Jakowatz, C.V.},  journal={IEEE Transactions on Aerospace and Electronic Systems},   title={Phase gradient autofocus-a robust tool for high resolution {SAR} phase correction},   year={1994},  volume={30},  number={3},  pages={827-835},  doi={10.1109/7.303752}}

@ARTICLE{InISAR_Zhang2004,
	author={Qun Zhang and Tat Soon Yeo and Gan Du and Shouhong Zhang},
	journal={IEEE Transactions on Geoscience and Remote Sensing}, 
	title={{Estimation of three-dimensional motion parameters in interferometric ISAR imaging}}, 
	year={2004},
	volume={42},
	number={2},
	pages={292-300},
	doi={10.1109/TGRS.2003.815669}}

@article{Javadi2022,
	author = {Javadi, S. Hamed and Hichem Sahli and Andr\'{e} Bourdoux},
	journal = {Appl. Opt.},
	number = {17},
	pages = {F1--F7},
	publisher = {Optica Publishing Group},
	title = {{Rayleigh-based segmentation of ISAR images}},
	volume = {62},
	month = {Jun},
	year = {2023},
	doi = {10.1364/AO.482527},
}

@techreport{Doerry_PFA,
	author      = {Doerry, Armin W.},
	title       = {{Basics of Polar-Format Algorithm for Processing Synthetic Aperture Radar Images}},
	institution = {Sandia National Laboratories},
	year        = {2012},
	number      = {SAND2012-3369},
	month       = {May},
}

@techreport{Doerry_notes,
	author      = {Doerry, Armin W. and Bickel, D. L.},
	title       = {{Notes on Synthetic Aperture Radar Image Quality}},
	institution = {Sandia National Laboratories},
	year        = {2024},
	number      = {SAND2024-16813},
	month       = {December},
}

@techreport{Doerry_BPA,
	author      = {Armin W. Doerry, Edward E. Bishop, John A. Miller},
	title       = {{Basics of Backprojection Algorithm for Processing Synthetic Aperture Radar Images}},
	institution = {Sandia National Laboratories},
	year        = {2016},
	number      = {SAND2012-3369},
	month       = {Feb.},
}

@book{ghiglia1998two,
	title={Two-Dimensional Phase Unwrapping: Theory, Algorithms, and Software},
	author={Ghiglia, Dennis C. and Pritt, Mark D.},
	year={1998},
	publisher={Wiley},
	address={New York},
	isbn={978-0-471-24935-1}
}

@book{Richards2014fundamentals,
	title={Fundamentals of Radar Signal Processing},
	author={Richards, Mark A.},
	year={2014},
	edition={2nd},
	publisher={McGraw-Hill Education}
}

@ARTICLE{mea2014,
	author={Xiong, Tao and Xing, Mengdao and Wang, Yong and Wang, Shuang and Sheng, Jialian and Guo, Liang},
	journal={IEEE Transactions on Geoscience and Remote Sensing}, 
	title={{Minimum-Entropy-Based Autofocus Algorithm for SAR Data Using Chebyshev Approximation and Method of Series Reversion, and Its Implementation in a Data Processor}}, 
	year={2014},
	volume={52},
	number={3},
	pages={1719-1728},
	doi={10.1109/TGRS.2013.2253781}}

@ARTICLE{Bamlery_1998,
  		author={Richard Bamlery and Philipp Hartl},
  		journal={Inverse Problems}, 
  		title={Synthetic Aperture Radar Interferometry}, 
  		year={1998},
  		volume={14},
  		number={},
  		pages={R1—R54}}

@INPROCEEDINGS{Javadi_altimetry_radarConf2025,
  		author={Javadi, S. Hamed and Bourdoux, André and Cappelle, Hans and Sahli, Hichem},
  		booktitle={2025 IEEE Radar Conference (RadarConf25)}, 
  		title={{Radar-Based Altimetry and Nadir Depth Profile Estimation for UAVs}}, 
  		year={2025},
  		volume={},
  		number={},
  		pages={1439-1444},
  		doi={10.1109/RadarConf2559087.2025.11205136}}

@INPROCEEDINGS{Javadi_depth_map_eurad2025,
  		author={Javadi, S. Hamed and Bourdoux, André and Sahli, Hichem},
  		booktitle={2025 22nd European Radar Conference (EuRAD)}, 
  		title={{Depth map reconstruction from low-altitude UAV}}, 
  		year={2025},
  		volume={},
  		number={},
  		pages={75-78}}

@ARTICLE{Javadi_leca_2025,
  		author={Javadi, S. Hamed and Bourdoux, André and Albaba, Adnan and Sahli, Hichem},
  		journal={IEEE Transactions on Radar Systems}, 
  		title={{A Low-Complexity PFA-Based Autofocus Algorithm for Automotive SAR}}, 
  		year={2025},
  		volume={3},
  		number={},
  		pages={799-810},
  		doi={10.1109/TRS.2025.3574010}}

@Article{mbpu_lin_2022,
  		AUTHOR = {Lin, Zihao and Duan, Yan and Deng, Yunkai and Tian, Weiming and Zhao, Zheng},
  		TITLE = {{An Improved Multi-Baseline Phase Unwrapping Method for GB-InSAR}},
  		JOURNAL = {Remote Sensing},
  		VOLUME = {14},
  		YEAR = {2022},
  		NUMBER = {11},
  		ARTICLE-NUMBER = {2543},
  		DOI = {10.3390/rs14112543}
  	}

@ARTICLE{Manzoni2023,
  		author={Manzoni, Marco and Tagliaferri, Dario and Rizzi, Marco and Tebaldini, Stefano and Guarnieri, Andrea Virgilio Monti and Prati, Claudio Maria and Nicoli, Monica and Russo, Ivan and Duque, Sergi and Mazzucco, Christian and Spagnolini, Umberto},
  		journal={IEEE Transactions on Intelligent Transportation Systems}, 
  		title={{Motion Estimation and Compensation in Automotive MIMO SAR}}, 
  		year={2023},
  		volume={24},
  		number={2},
  		pages={1756-1772},
  		doi={10.1109/TITS.2022.3219542}}

@ARTICLE{Simard2016,
  		author={Simard, Marc and Riel, Bryan V. and Denbina, Michael and Hensley, Scott},
  		journal={IEEE Transactions on Geoscience and Remote Sensing}, 
  		title={Radiometric Correction of Airborne Radar Images Over Forested Terrain With Topography}, 
  		year={2016},
  		volume={54},
  		number={8},
  		pages={4488-4500},
  		doi={10.1109/TGRS.2016.2543142}}

@incollection{sarhandbook2019chapter5,
  		author    = {Simard, Marc and others},
  		title     = {{SAR Methods for Mapping and Monitoring Forest Degradation and Deforestation}},
  		booktitle = {SAR Handbook: Comprehensive Methodologies for Forest Monitoring and Biomass Estimation},
  		publisher = {NASA},
  		year      = {2019}
  	}

@ARTICLE{Brenner2008,
  		author={Brenner, Andreas R. and Roessing, Ludwig},
  		journal={IEEE Transactions on Geoscience and Remote Sensing}, 
  		title={{Radar Imaging of Urban Areas by Means of Very High-Resolution SAR and Interferometric SAR}}, 
  		year={2008},
  		volume={46},
  		number={10},
  		pages={2971-2982},
  		doi={10.1109/TGRS.2008.920911}}

@ARTICLE{Eineder2009,
  		author={Eineder, Michael and Adam, Nico and Bamler, Richard and Yague-Martinez, Nestor and Breit, Helko},
  		journal={IEEE Transactions on Geoscience and Remote Sensing}, 
  		title={{Spaceborne Spotlight SAR Interferometry With TerraSAR-X}}, 
  		year={2009},
  		volume={47},
  		number={5},
  		pages={1524-1535},
  		doi={10.1109/TGRS.2008.2004714}}

@ARTICLE{Richards_InSAR_2007,
  		author={Richards, Mark A.},
  		journal={IEEE Aerospace and Electronic Systems Magazine}, 
  		title={{A Beginner's Guide to Interferometric SAR Concepts and Signal Processing [AESS Tutorial IV]}}, 
  		year={2007},
  		volume={22},
  		number={9},
  		pages={5-29},
  		doi={10.1109/MAES.2007.4350281}}

@ARTICLE{uav_sar_Bekar_2022,
  		author={Bekar, Ali and Antoniou, Michail and Baker, Christopher J.},
  		journal={IEEE Transactions on Geoscience and Remote Sensing}, 
  		title={{Low-Cost, High-Resolution, Drone-Borne SAR Imaging}}, 
  		year={2022},
  		volume={60},
  		number={},
  		pages={1-11},
  		doi={10.1109/TGRS.2021.3085235}}

@article{Svedin_Bernland_Gustafsson_Claar_Luong_2021, title={{Small UAV-based SAR system using low-cost radar, position, and attitude sensors with onboard imaging capability}}, volume={13}, DOI={10.1017/S1759078721000416}, number={6}, journal={International Journal of Microwave and Wireless Technologies}, author={Svedin, Jan and Bernland, Anders and Gustafsson, Andreas and Claar, Eric and Luong, John}, year={2021}, pages={602–613}}

@INPROCEEDINGS{tomoSAR_Wang_2024,
  		author={Wang, Yuhan and Zhao, Jian and Wang, Zhen and Zhu, Kaiwen and Dong, Zehua and Ding, Zegang},
  		booktitle={2024 IEEE International Conference on Signal, Information and Data Processing (ICSIDP)}, 
  		title={{UAV-SAR Panoramic 3D Imaging for Urban Area: A Multi-angle Tomography Approach}}, 
  		year={2024},
  		volume={},
  		number={},
  		pages={1-5},

  		doi={10.1109/ICSIDP62679.2024.10868643}}

@ARTICLE{Lahmeri2025,
  		author={Lahmeri, Mohamed-Amine and Mustieles-Pérez, Víctor and Vossiek, Martin and Krieger, Gerhard and Schober, Robert},
  		journal={IEEE Transactions on Communications}, 
  		title={{UAV Formation and Resource Allocation Optimization for Communication-Assisted 3D InSAR Sensing}}, 
  		year={2025},
  		volume={73},
  		number={8},
  		pages={5788-5804},
  		doi={10.1109/TCOMM.2025.3535902}}

@ARTICLE{Kellner2019,
  		author={Kellner, JR and Armston, J. and Birrer, M. and Cushman, KC. and Duncanson, L. and Eck, C. and Falleger, C. and Imbach, B. and Král, K. and Krůček, M. and Trochta, J. and Vrška, T. and Zgraggen, C.},
  		journal={Surv Geophys.}, 
  		title={New Opportunities for Forest Remote Sensing Through Ultra-High-Density Drone Lidar}, 
  		year={2019},
  		volume={40},
  		number={4},
  		pages={959-077},
  		doi={10.1007/s10712-019-09529-9}}

@article{Klemas2015,
  		author = {Victor V. Klemas},
  		title = {{Coastal and Environmental Remote Sensing from Unmanned Aerial Vehicles: An Overview}},
  		volume = {31},
  		journal = {Journal of Coastal Research},
  		number = {5},
  		publisher = {Coastal Education and Research Foundation},
  		pages = {1260 -- 1267},
  		year = {2015},
  		doi = {10.2112/JCOASTRES-D-15-00005.1},
  		URL = {https://doi.org/10.2112/JCOASTRES-D-15-00005.1}
  	}

@article{Gano2024,
  		author = {Gano, Boubacar and Bhadra, Sourav and Vilbig, Justin M. and Ahmed, Nurzaman and Sagan, Vasit and Shakoor, Nadia},
  		title = {Drone-based imaging sensors, techniques, and applications in plant phenotyping for crop breeding: A comprehensive review},
  		journal = {The Plant Phenome Journal},
  		volume = {7},
  		number = {1},
  		pages = {e20100},
  		doi = {https://doi.org/10.1002/ppj2.20100},
  		year = {2024}
  	}

@ARTICLE{Drone_InSAR_2025,
  		author={Mustieles-Perez, Victor and Kim, Sumin and Kanz, Julian and Bonfert, Christina and Grathwohl, Alexander and Krieger, Gerhard and Villano, Michelangelo},
  		journal={IEEE Journal of Selected Topics in Applied Earth Observations and Remote Sensing}, 
  		title={{Generation of Accurate, High-Resolution Digital Elevation Models From Ultrawideband, Drone-Borne, Repeat-Pass Interferometric SAR}}, 
  		year={2025},
  		volume={18},
  		number={},
  		pages={28375-28392},
  		doi={10.1109/JSTARS.2025.3626140}}

@ARTICLE{Xing2009,
  		author={Xing, Mengdao and Jiang, Xiuwei and Wu, Renbiao and Zhou, Feng and Bao, Zheng},
  		journal={IEEE Transactions on Geoscience and Remote Sensing}, 
  		title={{Motion Compensation for UAV SAR Based on Raw Radar Data}}, 
  		year={2009},
  		volume={47},
  		number={8},
  		pages={2870-2883},

  		doi={10.1109/TGRS.2009.2015657}}

@ARTICLE{Grassi2026,
  		author={Grassi, Pietro and Manzoni, Marco and Tebaldini, Stefano and Maria Prati, Claudio},
  		journal={IEEE Transactions on Geoscience and Remote Sensing}, 
  		title={{A Geometrical Autofocus Method for UAV-Based SAR}}, 
  		year={2026},
  		volume={64},
  		number={},
  		pages={1-17},
  		doi={10.1109/TGRS.2026.3655139}}
\end{document}